\documentstyle[12pt,{user:[valle.texfile]caption}]{article}
\font\tenbf=cmbx10
\font\tenrm=cmr10
\font\tenit=cmti10
\font\elevenbf=cmbx10 scaled\magstep 1
\font\elevenrm=cmr10 scaled\magstep 1
 1

\renewenvironment{thebibliography}[1]
 { \elevenrm
   \begin{list}{\arabic{enumi}.}
    {\usecounter{enumi} \setlength{\parsep}{0pt}
     \setlength{\itemsep}{3pt} \settowidth{\labelwidth}{#1.}
     \sloppy
    }}{\end{list}}
\include{matex}
\begin{document}
\begin{center}{{\tenbf
               BEYOND THE STANDARD MODEL IN TAU DECAYS\\}
\vglue 1.0cm
{\tenrm JOS\'E W. F. VALLE}
\footnote{E-mail VALLE at vm.ci.uv.es or 16444::VALLE}\\
\baselineskip=13pt
{\tenit Instituto de F\'{\i}sica Corpuscular - C.S.I.C.\\
Departament de F\'isica Te\`orica, Universitat de Val\`encia\\}
\baselineskip=12pt
{\tenit 46100 Burjassot, Val\`encia, SPAIN         }\\
\vglue 0.8cm
{\tenrm ABSTRACT}}
\end{center}
\vglue 0.3cm
{\rightskip=3pc
 \leftskip=3pc
 \tenrm\baselineskip=12pt
 \noindent
The tau sector may be substantially different than
predicted in the standard model. It can have lepton
flavour violating (LFV) $\tau$ decays such as
	$\tau \ra e \pi^0$ and
	$\tau \ra e \gamma$.
If these decays are induced by singlet
neutral heavy lepton (NHL) exchange
they may occur at the $\O(10^{-6})$ level.
I also discuss two-body decays with the emission
of a superweakly interacting massless spin zero
particle, called majoron, \eg in $\tau \ra \mu + J$
whose branching ratios can be at the level of present
sensitivities ($\O(10^{-3})$) without violating any
experimental data.
The underlying physics may be probed also at LEP through
related single NHL or chargino production processes,
e.g. $Z \ra \Nt \nt$ or $Z \ra \chi \tau$
(\Nt is a NHL while $\chi$ denotes the lightest
chargino). Existing observations allow these rates
to be large even when the associated \nt mass is
too small to be probed with present or
expected sensitivities.
\vglue 0.6cm}
{\elevenbf\noindent Introduction}
\hspace{\parindent}

Although our present description of particle
physics via the standard \21 model has been
extremely successful, it leaves several open
puzzles that motivate further extensions.

One of the puzzles of the \sm is the apparent
absence of right-handed neutrinos in nature
indicated, e.g. by the experimental limits on
neutrino masses and mixing. Theoretically,
there is no fundamental symmetry that
dictates the masslessness of neutrinos and it may,
in fact, be in conflict with cosmological and
astrophysical observations involving, for example,
dark matter and solar neutrinos. The former would
indicate \neu masses at a scale $\sim 10$ eV while
the latter suggest very small neutrino masses
$\sim 10^{-3}$ eV, needed to explain the solar
neutrino data via the MSW effect. One attractive
way to generate such \neu masses is through the
exchange of singlet neutral heavy leptons (NHLS)
or as radiative effects of new scalar bosons
\cite{fae}.

Another fundamental problem in electroweak physics
today is that of mass generation, which relies
on the Higgs mechanism and, in turn, implies
the existence of fundamental scalar bosons \cite{HIGGS}.
It is widely believed that some stabilizing principle
- e.g. supersymmetry (SUSY) - should be effective at
the electroweak scale in order to explain the stability
of this scale against quantum corrections related to
physics at superhigh scales. Unfortunately there is no
clue as to how SUSY is realized. The most popular $ansatz$
- called the minimal supersymmetric standard model (MSSM)
\cite{mssm} - realizes SUSY in the presence of a discrete
R parity ($R_p$) symmetry under which all standard model
particles are even while their partners are odd.
In this ansatz neutrinos are massless, as in the
standard model. However, this choice has no
firm theoretical basis and there are interesting
SUSY theories without R parity \cite{fae}.
Here we focus on the case of spontaneous $R_p$
breaking in the \21 theory (RPSUSY), where the
breaking of R-parity is driven by {\sl isosinglet}
slepton vacuum expectation values (VEVS) \cite{MASI,pot3}.
In this case the associated Goldstone boson (majoron)
is mostly singlet and as a result the $Z$ does not
decay by majoron emission, in agreement with LEP
observations \cite{LEP1}.

There is a wealth of phenomena associated with
massive neutrinos or RPSUSY \cite{granada}.
Either way of extending the
\sm leads to rare tau decays \cite{fae}.
These are closely related with exotic tau properties, such as
lepton universality violating couplings and/or nonzero \nt mass.
In this talk I focus on these $\tau$ number violating decays,
and their possible relation with the single production of
SUSY particles and/or the existence of new particles (new
scalar bosons or NHLS). In the latter case I stress that
lepton flavour violating processes may occur with sizeable
rates even when \neus are strictly massless or
extremely light. These rare tau decays
would be accompanied by rare $Z$ decays that
could also be accessible to experiment,
as I will illustrate with examples.

\vglue 0.6cm
{\elevenbf\noindent Preliminaries}
\hspace{\parindent}

There are several bounds on \neu masses that follow
from observation. The laboratory bounds may be
summarized as \cite{PDG92}
\beq
m_{\nu_e} 	\lsim 9 \: \rm{eV}, \:\:\:\:\:
m_{\nu_\mu}	\lsim 270 \: \rm{keV}, \:\:\:\:\:
m_{\nu_\tau}	\lsim 31  \: \rm{MeV}
\eeq
In addition there are limits on neutrino mass
and mixing that follow from the nonobservation of
neutrino oscillations, which I will not repeat
here \cite{PDG92}.

Apart fom laboratory limits, there is a cosmological limit that
follows from considerations related to the abundance
of relic neutrinos \cite{KT}
\beq
\sum_i m_{\nu_i} \lsim 50 \: eV
\label{rho1}
\eeq
This limit only holds if \neus are stable.
Indeed, there are many ways to make \neus unstable
in such a way as to avoid the limit in \eq{rho1}
\cite{fae}. The models rely on the existence
of fast \neu decays involving majoron emission
\cite{V,GV,CON,ROMA}, \eg
\beq
\nu_\tau \ra \nu_\mu + J
\label{NUJ}
\eeq
The resulting lifetime can be made sufficiently
short so that neutrino mass values as large as
present laboratory limits are fully consistent
with astrophysics and cosmology. Examples of
seesaw type models where this is possible
have been discussed in ref. \cite{GV,CON}
and a typical lifetime versus mass relationship
has been given in ref. \cite{CON}. Another
example is provided by the spontaneously
broken R parity (RPSUSY) model \cite{ROMA},
for which the minimum attainable \nt lifetimes
consistent with observation are given in Fig. 1,
as a function of the \nt mass.
They should be compared to the cosmological limit on
the \nt decay lifetime required in order to efficiently
suppress the relic \nt contribution. This is shown as the
solid straight line in Fig. 1. Clearly the decay lifetimes
can be shorter than required by cosmology. Moreover, since
these decays are $invisible$, they are consistent with all
astrophysical observations \cite{KT}. If, however, the
universe is to have become matter-dominated by a redshift
of 1000 at the latest (so that fluctuations have grown by
the same factor by today), the \nt lifetime has
to be much shorter \cite{ST}, as indicated by the dashed
line in Fig. 1. Again, lifetimes below the dashed line are
possible. However, this lifetime limit is less reliable
than the one derived from the critical density, since
there is not yet an established theory for the formation
of structure in the universe.
\bef
\vspace{9cm}
\label{ntdecay}
\caption{
Estimated \nt lifetime versus observational limits}
\eef

In addition to limits, observation also
provides some positive hints for neutrino masses.
These follow from cosmological, astrophysical
and laboratory observations which we now discuss.

Recent observations from COBE indicate
the existence of hot dark matter, for
which the most attarctive candidate is
a massive \neu, \sa a stable \nt,
with mass larger than a few eV \cite{KT}.
This suggests the possibility of having
observable \ne or \nm - \nt oscillations in the
laboratory. With good luck the next generation
of experiments such as CHORUS and NOMAD at CERN
and the P803 experiment proposed at Fermilab
will probe this possibility. In addition to
\neu oscillation signatures, some models also
suggest the possible existence of rare
muon or tau decays with appreciable rates
\cite{DARK}. The latter could be well probed
at a tau factory.

Second, the solar \neu data collected
up to now suggests the existence of \neu
conversions involving very small \neu masses
$\lsim 10^{-2}$ eV. The region of parameters
allowed by present experiments is illustrated
in Fig. 2 \cite{GALLEX}.
\bef
\vspace{13cm}
\label{msw}
\caption{Region of \neu oscillation parameters
allowed by experiment}
\eef
Here it is
interesting to remark that the recent results
of the GALLEX experiment on low energy pp \neus do not
really "eliminate" the solar \neu puzzle,
in view of the persisting deficit of high energy
\neus seen in Kamiokande and Homestake. The
astrophysical explanation of the latter data would
require not only too large a drop in the
temperature of the solar core, but also
would predict wrongly the relative degree of
suppression observed in these two experiments
\cite{Smirnov_wein}.

It is worth noting that there are some hints,
albeit controversial, from recent beta decay
studies based on solid state detectors which
indicate the presence of a 17 keV component
\cite{norman_wein}. This would require the
existence of a decay of the type in \eq{NUJ}.
The importance that such an observation would
have justifies the effort necessary to obtain
a conclusive confirmation or refutal of this result.

Finally, there are hints from studies involving
atmospheric neutrinos, which indicate the existence
a muon deficit \cite{atm}, etc. However I will not
discuss these in this talk.

\vglue 0.6cm
{\elevenbf\noindent Rare Tau Decays and NHLS}
\hspace{\parindent}

Neutral isosinglet heavy leptons (NHLS) arise in many
extensions of the electroweak theory \cite{SST}.
They can engender decays that are exactly forbidden
in the standard model, and whose detection would
signal new physics, closely related with the properties
of the \neus and the leptonic weak interaction \cite{fae}.
For this reason, one may argue that such processes, if nonzero,
ought to be very small. This is the typical situation
to expect when their existence is directly related to
nonzero \neu masses and therefore suppressed. However,
as discussed in \cite{BER,CP1,CP2,CERN} this need not
be the case. NHLS can mediate large LFV decays, such as
those shown in table 1.
\begin{table}
\begin{center}
\begin{math}
\begin{array}{|c|cr|} \hline
\rm channel & \rm strength & \\
\hline
Z \ra \Nt \nt &  \sim 10^{-3} & \\
Z \ra e \tau &  few \times 10^{-7} & \\
Z \ra \mu \tau &  \sim 10^{-7} & \\
\hline
\tau \ra e \gamma ,\mu \gamma &  \sim 10^{-6} & \\
\tau \ra e \pi^0 ,\mu \pi^0 &  \sim 10^{-6} & \\
\tau \ra e \eta^0 ,\mu \eta^0 &  few \times 10^{-7} & \\
\tau \ra 3e , 3 \mu , \mu \mu e, \etc &  few \times 10^{-7} & \\
\hline
\end{array}
\end{math}
\end{center}
\caption{Attainable branching ratios for $Z$ and $\tau$
decays consistent with lepton universality. The underlying
models involve isosinglet neutral heavy leptons,
and the usual neutrinos \ne, \nm and \nt may be
strictly massless.}
\end{table}
As shown in Fig. 3 (taken from ref. \cite{3E})
present constraints on weak universality violation
allow decay branching ratios as large as ${\cal O} (10^{-6})$.
The solid line corresponds to the attainable branching for the
decay $l \ra l_i l_j^+l_j^-$, the dashed line
corresponds to $\tau \ra \pi^0 l_i$, the dotted line
to $\tau \ra \eta l_i$ and the dash-dotted line to
$\tau \ra l_i \gamma$. Here $l_i$ denotes $e$ or $\mu$
and all possible final-state leptons have been summed
over in each case. As one can see, the most favorable
of all the $\tau$ decay channels are $\tau \ra e \gamma$
and $\tau \ra e \pi^0$ \cite{3E}, the first being dominant
for lower NHL masses in the $100\: \rm GeV - 10 \: \rm TeV$
range.
\bef
\vspace{6cm}
\caption{Attainable branching ratios for LFV $\tau$ decays
consistent with lepton universality. These processes
may occur even if \neus are strictly massless.}
\eef

Any novel physics, such as discussed here for
the $\tau$ lepton is bound to be manifest, one
way or the other, also at the $Z$ pole, and thus
produce signals at LEP. The search for rare $Z$
decays gives another way to test fundamental
symmetries of the standard electroweak theory,
such as weak universality, flavour and CP
conservation. The underlying physics is the same,
although the detectability prospects may differ.
For example, NHLS can induce new $Z$ decays,
such as
\footnote{There may also be CP asymmetries in these decays,
despite the fact that the physical light \neus are strictly
massless \cite{CP1}. These asymmetries can be of order unity
with respect to the corresponding LFV decays \cite{CP2}. }
\beq
\begin{array}{lr}
Z \ra e + \tau \: , & Z \ra N_{\tau} + \nu_{\tau}
\end{array}
\eeq
Taking into account the constraints on the parameters
describing the leptonic weak interaction one can
estimate the attainable values for these branching
ratios given in table 1. The most copious channel
is $Z \ra \Nt \nt$, possible if the \Nt is lighter
than the $Z$ \cite{CERN}. Subsequent \Nt decays
would then give rise to large missing $p_T$ events,
called zen-events. Prompted by our suggestion,
there are now good limits on such decays from the
searches for acoplanar jets and leptons from $Z$
decays at LEP \cite{opal}. The method of
experimental analysis is very similar to that
for SUSY zen events.

Under realistic assumptions,
one is a bit too short to be able to see $Z \ra e \tau$
or $Z \ra \mu \tau$ at LEP, as can be seen from Fig. 3.22
and 3.23 of ref. \cite{ETAU}. In contrast, the related
low energy processes seem to be within the expectations
of a $\tau$ or $B$ factory. In any case, there have
been dedicated searches for flavour violation at the
$Z$ peak at LEP, and some limits have already been
obtained \cite{opal}.
This example illustrates how this physics is
{\it complementary to the physics of
\neu mass per se}. For more details the reader
is referred to ref. \cite{BER,CP1,CP2,CERN,3E}.

\vglue 0.6cm
{\elevenbf\noindent Rare Tau Decays and Supersymmetry}
\hspace{\parindent}

Supersymmetris models can produce rare $Z$ and $\tau$ decays
with detectable rates.  Here we are concerned with the
situation where supersymmetry is realized in a \21 context
in such a way that R parity is broken spontaneously at or
slightly below the TeV scale \cite{NPBTAU,MASI,ROMA}.
In such RPSUSY models one can have single SUSY particle
production processes, such as
\beq
\begin{array}{lr}
Z \ra \chi + \tau \: , 
\end{array}
\eeq
The single production of the lightest $chargino$
would lead to striking signatures at LEP \cite{ROMA}.
The allowed branching ratio lies close to the
present LEP sensitivities, as shown in Fig. 4.
\bef
\vspace{6cm}
\label{}
\caption{Attainable branching ratios for single
chargino production in $Z$ decays, $Z \ra \chi + \tau$,
as a function of the chargino and \nt masses.
Comparing this with Fig. xx illustrates the complementarity
of $Z$ and $\tau$ physics.}
\eef
In addition, in this model the lightest neutralino
is unstable and is therefore not necessarily
an origin of events with missing energy.
Zen events, similar to those of the
MSSM are expected say, from
the single neutralino production process, where
$\chi^0$ decays visibly and the missing energy
is carried by the \nt. The corresponding zen-
event rates can be larger than in the minimal
SUSY model and their origin is also quite different.
Moreover, these processes are intimately tied with
each other and to the nonzero value of the \nt mass,
as shown in Fig. 2.

Since in this model also lepton number is
broken spontaneously there is a physical Goldstone
boson, called Majoron. Its existence is consistent with
recent measurements of the $invisible$ $Z$ decay width
at LEP, since the Majoron is a singlet under the \21
\gau symmetry.

My second example of LFV $\tau$ decays involves
precisely the decay to light scalar particles,
such as the majoron. Such weakly interacting
majoron would be "seen" only insofar as it
would affect the spectra of the leptons
produced in $\tau$ decays.
Single majoron emission in $\tau$ decays would lead
to bumps in the final lepton energy spectrum,
at half of the $\tau$ mass in its rest frame.
The allowed rates have been determined by varying
the relevant parameters over reasonable ranges
and imposing all of the observational constraints
(for details, see ref. \cite{NPBTAU}).
A specially important role is played by
constraints related to the flavour and/or
total lepton number violating processes \sa
those arising from negative neutrino oscillation
and neutrinoless double $\beta$ decay searches,
as well as from the failure to observe
anomalous peaks on the energy distribution of
the electrons and muons coming from decays such
as $\pi, K \ra e \nu$ and $\pi, K \ra \mu \nu$.
Our results are shown in Fig. 3, taken from ref.
\cite{NPBTAU}. One sees that these modes
may occur at a level ${\cal O} (10^{-3})$ in branching ratio
\cite{NPBTAU}, not so far from existing limits \cite{SINGLE}.
\bef
\vspace{6cm}
\caption{Attainable branching ratios for LFV $\tau$
decays with Majoron emission, as a function of the
\nt mass.}
\eef
One sees that the Majoron emission $\tau$ decay branching
ratio can easily lie in the interesting range between
$10^{-5}$ and the present ARGUS limit \cite{SINGLE}
$BR(\tau \ra \mu + J) \leq 5.8 \times 10^{-3}$.
The points in the region delimited by the solid contours
correspond to different values of $\tan\beta$, an
unknown model parameter expected to lie between 1
and $\frac{m_t}{m_b}$, the top-bottom quark mass ratio.
Similar results hold for the case of the decay $ \tau \ra e + J$.
In this case the attainable branching ratio cannot exceed
$\rm few \times 10^{-4}$, for $\beta$ $\tan\beta \leq 40$.
This is to be contrasted with the present ARGUS limit
$BR(\tau \ra e + J) \leq 3.2 \times 10^{-3}$
(see Fig. 4 of ref. \cite{NPBTAU}).

\vglue 0.6cm
{\elevenbf\noindent Discussion}
\hspace{\parindent}

Upcoming tau factories, or even a $B$ factory,
will be ideally suited to probe for the possible
existence of the $\tau$ decay processes discussed
here. If no signal is found, one will obtain
restrictions on the relevant parameters of
these extensions of the \smp
The physics of a $Z$ factory beautifully complements
what can be learned from a {\it flavour factory}
where $\tau$ leptons can be copiously produced.
It may even become possible to do meaningful
rare tau decay searches using the taus produced at
LEP itself.
Finally, even if an improvement on the \nt mass
or universality limits is obtained in the near future,
it is not likely to narrow down the rare $Z$ and $\tau$
decay possibilities discussed here to a level where
they are rendered undetectable, both at LEP and/or
at a Tau Factory.
\vskip 5mm
This work was supported by CICYT-AEN-90-0040.
I thank J. Rom\~{a}o for preparing Fig. xx

\bibliographystyle{ansrt}

\end{document}